\documentclass[dvips,floatfix,intlimits,pra,showkeys,showpacs,twocolumn]{revtex4}

\usepackage{amsfonts,amsmath,amssymb}
\usepackage{dcolumn}
\usepackage{graphicx}
\usepackage{hyperref}

\begin{document}

\title{Loading Stark-decelerated molecules into electrostatic quadrupole traps}%
\author{Joop~J.~Gilijamse}%
\author{Steven~Hoekstra}%
\author{Nicolas~Vanhaecke}%
\thanks{Present address: Laboratoire Aim\'e Cotton, CNRS,
   B\^at 505, Universit\'e Paris-Sud, 91405 Orsay, France.}%
\author{Sebastiaan~Y.~T.~van~de~Meerakker}%
\author{Gerard~Meijer}%
\email[Author to whom correspondence should be sent. Electronic
mail:~]{meijer@fhi-berlin.mpg.de}%
\affiliation{Fritz-Haber-Institut der Max-Planck-Gesellschaft,
Faradayweg 4-6,
   14195 Berlin, Germany}%
\date{\today}%
\begin{abstract}\noindent%
Beams of neutral polar molecules in a low-field seeking quantum state can be
slowed down using a Stark decelerator, and can subsequently be loaded and
confined in electrostatic quadrupole traps. The efficiency of the trap loading
process is determined by the ability to couple the decelerated packet of molecules
into the trap without loss of molecules and without heating. We discuss the
inherent difficulties to obtain ideal trap loading, and describe and compare
different trap loading strategies. A new "split-endcap" quadrupole trap design is
presented that enables improved trap loading efficiencies. This is experimentally
verified by comparing the trapping of OH radicals using the conventional and the
new quadrupole trap designs.
\end{abstract}%
\keywords{cold molecules; Stark deceleration; trapping; phase-space dynamics}%
\pacs{}
\maketitle%

\section{Introduction}
\label{sec:introduction}

The ability to manipulate and control both the internal (rotation,
vibration) and external (velocity, orientation) degrees of freedom
of neutral molecules has lead to an increased interest in
gas-phase molecular physics. Several books and special issues of
journals have recently appeared in which the intense ongoing
research efforts in the rapidly emerging field of Cold Molecules
are documented
\cite{FaradayDisc142,Smith:LowTemperatures,Krems:ColdMolecules,NewJPhys:SpecialIssue}.
In these references, and in the original literature cited therein,
the various experimental routes to produce samples of trapped
neutral molecules together with their possible applications and
their anticipated or predicted properties are described in detail.
One of the experimental approaches is to start with a molecular
beam containing internally cold but fast moving neutral molecules,
and to then use (a combination of) electric, magnetic or radiative
fields to bring these molecules to a standstill and to confine
them in a trap \cite{Meerakker:NatPhys4:595}.

Deceleration of a beam of neutral molecules was first demonstrated on metastable
CO molecules using an array of electric fields in a so-called Stark decelerator
\cite{Bethlem:PRL83:1558}. The decelerated molecules can subsequently be
loaded and confined in a variety of traps. Electric field traps with a quadrupole
geometry, originally proposed by Wing for Rydberg atoms \cite{Wing:PRL45:631},
offer the steepest and deepest confining potentials, and have resulted, for
instance, in the trapping of ND$_3$ molecules \cite{Bethlem:Nature406:491}
and OH \cite{Meerakker:PRL94:023004} radicals. Traps with other field
geometries have been developed and tested as well. A four-electrode trap
geometry that combines a dipole, quadrupole and hexapole field has been
tested using decelerated ND$_3$ molecules \cite{Veldhoven:PRA73:063408}.
Confinement of Stark-decelerated OH radicals in combined magnetic and electric
fields \cite{Sawyer:PRL98:253002}, as well in a magnetic trap consisting of
rare-earth magnets \cite{Sawyer:PRL101:203203}, has recently been
demonstrated. With the Stark deceleration and trapping technique, samples of
cold molecules that are ideally suited for the measurement of the properties
of individual molecules, like lifetimes of metastable states, are now routinely produced
\cite{Meerakker:PRL95:013003,Gilijamse:JCP127:221102,Hoekstra:PRL98:133001}.
At present, the densities in the trap are not high enough to study collective
effects, however, as the densities are still too low for collisions to
occur between the trapped molecules on the timescale of the trap lifetime. Higher
number densities are required if one wants to apply evaporative or sympathetic
cooling to increase the phase-space density of the sample of trapped molecules,
and optimizing the density of trapped molecules therefore remains an important goal.

To reach the highest possible densities of molecules in the trap,
the deceleration and trap-loading process should be performed with
the lowest possible losses. During the deceleration process, low
losses are assured by the concept of phase stability; the motion
of the molecules through the decelerator is as if they were
trapped in a travelling potential well \cite{Bethlem:PRL84:5744}.
Although coupling of the transverse and the longitudinal motion in
the decelerator can lead to loss of molecules
\cite{Meerakker:PRA73:023401,Sawyer:EPJD48:197}, these losses can
be completely avoided when an improved mode of operation of the
decelerator is used \cite{Scharfenberg:PRA79:023410}. The concept
of phase-stability no longer holds when the molecules become too
slow, in particular at velocities that are required in the
trapping and trap loading region. The coupling of the decelerated
packet into the trap is therefore usually accompanied by
significant losses; the packet of slow molecules is not kept
together sufficiently well in six-dimensional phase-space upon
entering the trap. In this paper, the origin of the difficulty to
obtain efficient trap loading after Stark deceleration is
described in detail. We compare the different strategies that have
been implemented in our laboratory in recent years to optimize the
loading of Stark decelerated molecules in electrostatic quadrupole
traps. We present a new "split-endcap" quadrupole trap design that
reduces the losses during trap loading, without affecting the
shape and depth of the trapping potential. The improved
trap-loading is experimentally demonstrated by comparing the
signal of OH radicals in the split-endcap quadrupole trap, with
that in a conventional quadrupole trap.

\section{Loading of electrostatic quadrupole traps} \label{sec:new:criteria}

A trap can only confine molecules that have a position and velocity that are within
the so-called acceptance of the trap. The position and velocity distribution of the
beam that exits the decelerator is called the beam emittance of the decelerator.
In the ideal trap design, the (6D) beam emittance of the Stark decelerator is perfectly
mapped onto the (6D) acceptance of the trap \cite{Bethlem:PRA65:053416}. In
general, the acceptance of the trap can differ in size and shape from the emittance
of the decelerator. The shape of the longitudinal acceptance in a quadrupole trap
is identical to that of the transverse acceptance, and there is only a geometric
factor relating the size of the two. For the packet of slow molecules that exits the
decelerator, however, the beam emittance can be rather different in the longitudinal
and each of the transverse directions. In principle, a good 6D match can be achieved
by installing appropriate focusing elements and free flight sections between the
decelerator and the trap that allow for independent control over the longitudinal and
transverse motion of the molecules. A pulsed hexapole can be used, for instance, to
image the transverse phase-space distribution of the decelerated beam onto
the transverse trap acceptance. In the longitudinal direction, this can be achieved
when a buncher is used \cite{Crompvoets:PRL89:093004}. Although these elements
would allow for a good phase-space matching, in practice there are also disadvantages
to this approach. Most importantly, these elements significantly increase the distance
between the decelerator and trap. In view of the low longitudinal velocity that is required
for trap loading (typically 20~m/s or less at the exit of the last deceleration stage, just
before the loading), the packet of molecules will spatially expand significantly in the free
flight sections between the elements. This will inevitably lead to a large loss of molecules
as only a limited part of the beam can be manipulated and loaded into the trap. This
was indeed observed in the initial trapping experiments with ND$_3$ molecules where
a short hexapole~\cite{Bethlem:Nature406:491} or a bunching
element~\cite{Bethlem:PRA65:053416} were installed between the
decelerator and trap to allow for some phase-space manipulation of the
decelerated packet upon trap loading. In more recent trapping experiments with OH
radicals, the more pragmatic approach to install the trap as close as possible
to the exit of the decelerator was followed~\cite{Meerakker:ARPC57:159}. Although
the possibility to influence the phase-space distribution of the packet is compromised,
this strategy reduced the losses during the trap loading significantly. The problems
inherent to efficient trap loading have already been discussed in the first trapping
experiments of OH radicals~\cite{Meerakker:PRL94:023004}. It was observed then
that the signal of the trapped OH molecules was optimal when a packet of molecules
was loaded into the trap with a velocity that is actually too high. The packet spreads
out less upon entering the trap, but only comes to a standstill past the center of the trap.
This leads to a donut-shape longitudinal phase-space distribution that basically fills
the entire trapping volume. The impossibility to optimize simultaneously the number
and temperature of the trapped molecules in this trap design was also discussed in
an experiment where the decelerator and trap were optimized using evolutionary
strategies \cite{Gilijamse:PRA73:063410}.

\section{Motion of molecules during trap loading}

To appreciate the problems associated with the efficient loading of Stark-decelerated
molecules into an electrostatic trap, a more quantitative discussion of the trap loading
procedure is required. In this section, a number of trajectory calculations is presented
that illustrate the problems that can occur during the trap loading process. Different trap
loading procedures are discussed,  with emphasis at first on the evolution of the longitudinal
phase-space distribution of the packet of molecules during the trap loading process.
Trap loading procedures that can be used with the conventional quadrupole trap are
discussed in section~\ref{sec:new:old-loading}; the new split-endcap quadrupole trap
design is presented in section~\ref{sec:new:new-loading}. An experimental comparison
of the efficiency of these different trap loading strategies is presented in
section~\ref{sec:new:experiment}. In section~\ref{sec:transverse}, a discussion of the
evolution of the transverse phase-space distribution upon trap loading is given.

\subsection{Conventional quadrupole trap}
\label{sec:new:old-loading} 
\begin{figure}[htb!]
   \centering
   \resizebox{\linewidth}{!}
   {\includegraphics[0,0][364,250]{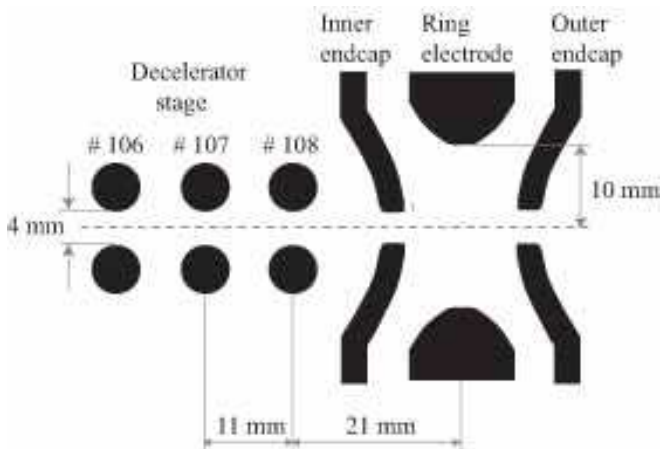}}
   \caption{Schematic representation of the geometry of the last stages of the decelerator
   and the trap region. The conventional electrostatic quadrupole trap consists of two hyperbolic endcaps and a ring electrode.}
   \label{fig:new:basicsetup}
\end{figure}

The geometrical details of the
conventional quadrupole trap that has been used so far in our
experiments are depicted in figure~\ref{fig:new:basicsetup}, and
have been described in detail elsewhere
\cite{Meerakker:ARPC57:159}. The figure shows a cut of the
cylindrically symmetric trap, consisting of a ring electrode and
two parabolic endcaps. The ring electrode is centered 21~mm
downstream from the last electrodes of the decelerator, and has an
inner radius~$R$ of 10~mm. The two hyperbolic endcaps have a
half-spacing of~$R/\sqrt{2}$. The 4~ and 6~mm diameter openings in
the left and right endcap allow for the entrance of the molecules
and for the outcoupling of the fluorescence light, respectively.
The last three electrode pairs of the Stark decelerator are shown
schematically; in reality they are placed alternatingly in
orthogonal transverse directions.

\begin{figure}[htb!]
   \centering
   \resizebox{\linewidth}{!}
   {\includegraphics[0,0][567,344]{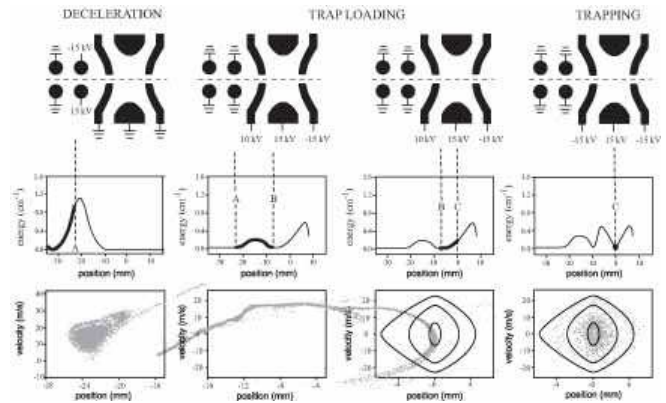}}
   \caption{Voltage configurations (top), corresponding potential
   energy curves for OH molecules along the beam axis (middle), and simulated longitudinal phase-space
   distributions (bottom) at various stages of the trap loading procedure.}
   \label{fig:new:old10kV}
\end{figure}

A pictorial presentation of the principle of the trap loading
process is shown in figure~\ref{fig:new:old10kV}. The parameters
that are used in this figure apply to a typical OH trapping
experiment, and are for a Stark decelerator and trapping machine
that is operational in our laboratory \cite{Meerakker:ARPC57:159}.
The voltages applied to the individual electrodes in the various
stages of the loading sequence are shown in the top of this
figure, and the corresponding potential energy curves for an OH
molecule that travels along the beam axis are shown directly
underneath. The part of the potential that the synchronous
molecule \cite{Bethlem:PRL84:5744} experiences during the loading
process is indicated by the thick lines. In the loading
configuration a voltage of 10~kV, 15~kV and -15~kV is applied to
the left endcap, ring electrode and right endcap, respectively.
This creates a quadratic loading potential that allows a maximum
velocity of 14.9~m/s for the OH radicals at the exit of the
decelerator in order to come to a standstill at the trap center.
The last stage of the decelerator switches off when the
synchronous molecule has reached position~A. At this time, the
electrodes are switched into the so-called \emph{loading}
configuration. Molecules experience the potential corresponding to
this configuration while the synchronous molecule moves from
position~A, via position~B, to position~C. When the synchronous
molecule has finally come to a standstill at position~C, the trap
center, the voltages on the electrodes are switched to the
\emph{trapping} configuration.
In the lower panel of figure~\ref{fig:new:old10kV}, the evolution
of the longitudinal phase-space distribution of the decelerated
packet is shown, as it results from numerical simulations of the
trap loading process. The distributions are shown at the times
at which the synchronous molecule has reached the positions~A, ~B
and~C. These times are also referred to as times~A, B and C. The
right-most panel shows the phase-space distribution of the cloud
after 20~ms of trapping.

At time~A the packet occupies a region in longitudinal phase-space
with a velocity spread of 7~m/s (FWHM), centered around a velocity
of 15~m/s. After the switching on of the loading potential, this
packet first has to overcome a small potential hill in the region
in between position~A and position~B. Since the velocity spread is
relatively large, a sizeable part of the molecules does
not possess enough kinetic energy to overcome this barrier and is
reflected, resulting in molecules with negative
velocity in the phase-space distribution at time~B (second graph
from the left in the lowest panel in figure~\ref{fig:new:old10kV}). From time~B
on, molecules are decelerated on the harmonic loading potential
and rotate in phase-space around the synchronous molecule. This
results in a rather poor overlap of the distribution of molecules with the trap
acceptance -- that is shown as an overlay -- at time~C. A large fraction
of the molecules is not within the trap acceptance, and
will not be confined when the trapping potential is switched on.

In the loading process as sketched above, a substantial fraction
of the molecules is reflected by the small "prebump" (between~A
and ~B) in front of the actual loading potential. In order to reduce
the losses associated with this reflection the voltage on the first
end-cap electrode in the loading configuration can be lowered.
This approach has indeed also been used in the past~\cite{Meerakker:PRL94:023004}
and has been demonstrated to improve the efficiency of the loading
process. A disadvantage of this approach, however, is that the
reduced voltage lowers the total height of the potential, and that a
slower packet of molecules is required to load the trap. In
addition, reduction of the voltage on the left endcap distorts the
harmonicity of the loading potential.

\begin{figure}[htb!]
   \centering
   \resizebox{\linewidth}{!}
   {\includegraphics[0,0][584,346]{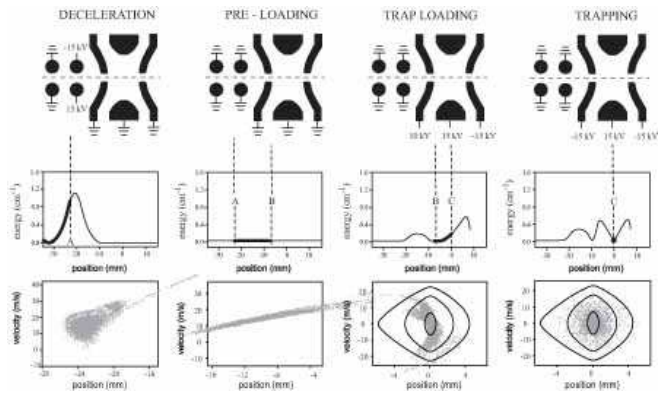}}
   \caption{Voltage configurations as applied in the loading sequence including a
   free flight section (top). Corresponding potential
   energy curves for OH molecules along the beam axis (middle).
   Simulated longitudinal phase-space distributions at various
   times during trap loading (bottom).}
   \label{fig:new:flatloading}
\end{figure}

An alternative approach to prevent losses due to reflection on the
prebump in front of the real loading potential, is to switch the
trap into the loading configuration only when the synchronous
molecule has reached position~B. In this case, there is no
prebump, and the molecular packet progresses in free flight to
the trap region. This procedure is sketched in
figure~\ref{fig:new:flatloading}. After the molecules are decelerated
to 15~m/s and exit the decelerator at time~A,
all voltages are switched off, leaving the packet in a field-free
region. At time~B the mean velocity of the packet is still 15~m/s,
but the distribution is stretched in position. At time~B, the loading configuration of the trap
is switched on and the molecules are decelerated to a mean
velocity of zero at time~C, where the molecules are subsequently
trapped.

\subsection{Split-endcap quadrupole trap}
\label{sec:new:new-loading}

\begin{figure}[htb!]
   \centering
   \includegraphics[width=\linewidth]{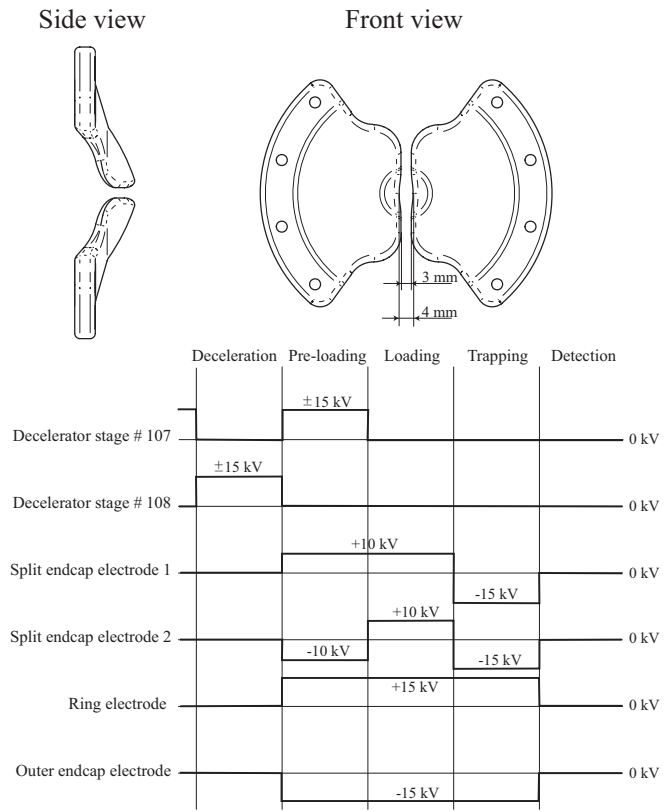}
   \caption{Side and front views of the split-endcap electrodes. The sequences of high voltage
   pulses that are applied to all four trap electrodes, as well as to the last two electrode
   pairs of the decelerator, are graphically shown.}
   \label{fig:new:split-endcap}
\end{figure}

In the trap loading strategies that are presented above, the trap
is positioned as close as possible to the exit of the Stark
decelerator. This is probably always preferred over designs that
include additional manipulation elements, but nevertheless the
trap loading is unsatisfactory. It appears that an improvement can
be obtained if the sequence of potentials that keeps the packet together
inside the decelerator can be extended into the trap region. Referring
back to figures~\ref{fig:new:old10kV} and~\ref{fig:new:flatloading},
the breakdown of this sequence is located between the positions~A
and~B. The basic idea of the split-endcap quadrupole trap design
presented here is to create a high potential hill in the region~AB that
can be used as an additional electric field stage of the decelerator,
effectively merging the Stark decelerator with the electrostatic
trap. It is noted that this idea has also been implemented in
trapping experiments using Stark-decelerated ND$_3$ molecules,
where the trap design that was used allowed for a straightforward
implementation of an additional electric field
stage~\cite{Schnell:JPCA111:7411}. For our cylindrically symmetric
quadrupole trap, this merging is achieved by breaking the symmetry
of the left endcap. In the new trap design this electrode is
replaced by two half endcaps with a small vacuum slit between the
two halves. The schematic side and front views of the split-endcap
electrodes, are shown in figure \ref{fig:new:split-endcap}.
A small curvature along the vacuum slit between the two halves of
the split-endcap is introduced to improve the transverse focusing
properties of the trap (see section~\ref{sec:transverse}).

\begin{figure}[htb!]
   \centering
   \resizebox{\linewidth}{!}
   {\includegraphics[0,0][700,375]{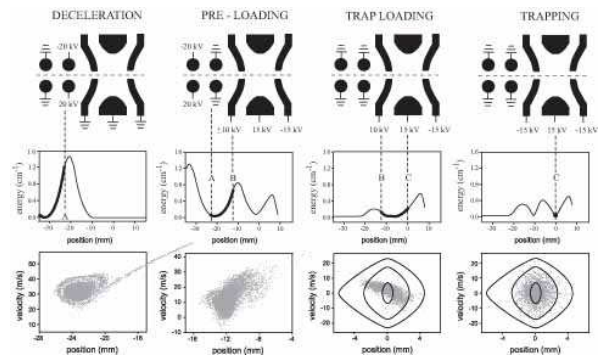}}
   \caption{The Stark decelerator and electrostatic trap are merged in the split-endcap
   electrode geometry. The potential hill in the region AB is used as an additional deceleration stage.}
   \label{fig:new:SECidea}
\end{figure}

Analogous to figure~\ref{fig:new:flatloading}, the voltages that
are applied to the electrodes, the on-axis potential energy curves
that the synchronous molecule experiences during the trap loading
process, and the evolution of the longitudinal phase-space
distribution of the packet of molecules, are shown in
figure~\ref{fig:new:SECidea}. For clarity, the sequence of high
voltage pulses that is applied to the last two decelerator stages
and to each of the four trap electrodes, are indicated in
figure~\ref{fig:new:split-endcap}.
As before, the last pair of electrodes of the decelerator is
switched to ground when the synchronous molecule has reached
position A. At this time, both halves of the split-endcap are
switched to high voltage (10~kV) of opposite parity, creating a
high electric field in between both split-endcap electrodes.
Simultaneously, the fore-last set of electrodes of the decelerator
are switched to $\pm$15~kV. In this so-called pre-loading
configuration, a potential hill in the region~AB is created that
is similar in shape to the series of potentials that are present
inside the Stark decelerator, and that can be used as an
additional deceleration stage. As a result, the forward velocity
of the packet of OH radicals at time~A can be higher (32~m/s, FWHM
6~m/s) than in the previously discussed trap loading schemes.
During this pre-loading configuration, the mean velocity of the
packet is reduced to 10~m/s (FWHM~12~m/s). It is evident from the
phase-space distribution, that the packet of molecules is kept
together much better during this period. When the synchronous
molecule has reached position B, the trap loading procedure
proceeds in the usual way. The trap can be used in the loading and
trapping configuration by switching both halves of the split
endcap to the same polarity, first to $+10$~kV and then to
$-15$~kV, respectively. The distance between the two electrodes is
small enough that the electric field distribution in the trap is
not significantly different from the original situation without a
gap, if both halves are switched to the same high voltage. During
the loading configuration, the remaining kinetic energy is taken
out on the loading slope such that the packet reaches zero
velocity in the center of the trap. Then, the trap is switched to
the trapping configuration, and the part of the molecular packet
that is within the acceptance of the trap stays confined.

It is seen that, compared to the trap loading strategies using the
cylindrically symmetric endcap, the longitudinal phase-space
distribution is spread out less at the moment the trap is switched
on. Even though the phase-space distribution is not perfectly matched
to the trap acceptance, almost all molecules are confined within
the innermost trap contours.

\subsection{Experiments}
\label{sec:new:experiment}

The different trap loading strategies presented above are
experimentally tested in a Stark deceleration molecular beam
machine. A detailed description of this machine, as well as of
the production and detection of OH radicals,
is given elsewhere \cite{Meerakker:ARPC57:159}. To experimentally
study the difference in trap loading efficiency between the
conventional and the split-endcap quadrupole trap, a direct
comparison between the old and new trap loading approaches is
required under otherwise identical conditions, i.e., without
removing and re-installing trap electrodes. For this, the original
cylindrically symmetric left endcap is replaced by the two
split-endcap electrodes. With the split-endcap quadrupole trap in
place, trap loading measurements that resemble the old trap
loading approach are still possible when the same high
voltage pulses are applied to both halves of the split-endcap.
Both split-endcap electrodes, the ring
electrode and the outer endcap electrode are individually
suspended and connected to an own set of high voltage switches. To
enable the switching between voltages of different amplitude
and polarity (note that the latter is required for both split-endcap
electrodes) a number of high voltage switches are configured in series, i.e.,
the output of one switch is connected to one of the input ports of
the next switch.

\begin{figure}[htb!]
   \centering
   \includegraphics[width=\linewidth]{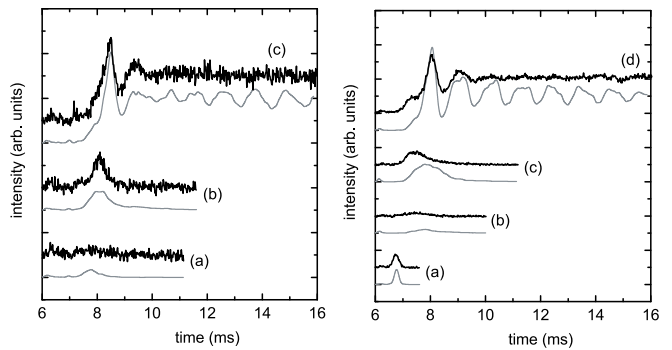}
   \caption{Time-of-flight profiles of OH radicals at different stages of
   the trap loading procedure that is depicted in figure~\ref{fig:new:flatloading} (left)
   and in figure~\ref{fig:new:SECidea} (right). The profiles that are obtained from
   three-dimensional trajectory simulations of the experiment are shown
   underneath the experimental profiles. The experimental and simulated profiles are
   given a vertical offset for clarity.}
   \label{fig:new:tofflatloading}
\end{figure}

For the three trap loading strategies, the loading of the
molecules into the trap is experimentally studied by terminating
the trap loading sequence at different stages of the trap loading
procedure, and by recording the time-of-flight profiles of the
slow packet of OH radicals at the center of the trap. In agreement
with earlier findings, the first strategy mentioned in
section~\ref{sec:new:old-loading} was found to be very inefficient
and resulted in a poor signal-to-noise ratio in the experiments.
These measurements are not shown here. The experimental results
that are obtained when the second loading strategy is followed, in
which the small potential barrier in the region AB is eliminated
by inserting a free-flight section
(figure~\ref{fig:new:flatloading}), are presented on the left-hand
side in figure~\ref{fig:new:tofflatloading}.
The Stark decelerator is programmed to produce a packet of OH
radicals with an average forward velocity of 15~m/s. In curve~(a)
the time-of-flight profile is shown of the packet that exits the
decelerator, i.e., the trap loading procedure is stopped after the
last stage of the decelerator is switched off. In this curve, the
signature of the slow packet of molecules is hardly visible. This
is due to the low forward velocity of 15~m/s (FWHM 6~m/s) of the
packet, and the spreading out of this packet while flying to the
trap center. In curve~(b) the time-of-flight profile is shown that
is observed when the "loading" part of the trap loading procedure
is included. The packet comes to a standstill in the center of the
trap, 8.0~ms after production. The higher signal intensity compared
to curve~(a) is due to the improved focusing properties when the
loading potential is present. Finally, in curve~(c), the full trap
loading sequence is used. After switching on of the trapping
potential, a large increase of signal followed by a damped
oscillation is observed. The time-of-flight profiles that result
from three dimensional trajectory simulations are shown underneath
the experimental profiles. Satisfactory agreement is obtained for
all curves, and the relative signal intensities of the different
profiles are reproduced rather well. The difference between the
simulated and measured curves regarding the damping of the
oscillation is discussed later in section~\ref{sec:transverse}.

On the right-hand side of figure~\ref{fig:new:tofflatloading} the measured
time-of-flight profiles are shown that correspond to the loading
strategy when the split-endcap electrodes are used as an
additional deceleration stage. The "deceleration",
"loading", and "trapping" profiles are shown in curves~(a),~(c),
and~(d), respectively. The additional profile that corresponds to
the "pre-loading" configuration, is shown in curve~(b).
The packet of molecules that exits the decelerator has a forward
velocity of 32~m/s with a velocity spread of 7~m/s, and arrives at
the trap center 6.8~ms after its production. Due to the higher
forward velocity, the packet spreads out less, and the signature
of the slow packet is clearly visible in the time-of-flight
profile. When the pre-loading configuration is added to the trap
loading sequence, curve~(b) is obtained. The packet arrives later
at the trap center, and the arrival time distribution is broader.
This reflects the lower forward velocity of 10~m/s. This very low
forward velocity also explains the reduced signal intensity; the
packet of molecules expands significantly when
the voltages on the trap electrodes are switched off in this
time-of-flight measurement. The packet is brought to a standstill
at the center of the trap when the loading part is added
(curve~(c)). The loading potential prevents the packet from
spreading out, and the signal intensity of the stopped molecules
is higher again. Curve~(d) corresponds to the full trap loading
sequence. Again, a large increase of signal followed by a damped
oscillation is observed when the trap is switched on. The
time-of-flight profiles that result from three dimensional
trajectory simulations are shown underneath the experimental
profiles. Again, satisfactory agreement is obtained for all
curves, although the slow packet in curves~(b) and (c) arrives
earlier in the experiment than in the simulation. This is
indicative of an over-estimate of the pre-loading potential in the
simulations. Deviations originate from misalignments of the trap
electrodes in the experiment, and from slight differences between the
actual and simulated shapes of the electrodes.

\subsection{Discussion}
\label{sec:new:Discussion}

The intensity of the fluorescence signal from the trapped sample
of molecules that is obtained when the three different trap
loading strategies are used can be directly compared. The
experimental results are summarized in
Table~\ref{tab:new:results}. In this Table, the three loading
strategies are referred to as "conventional loading", "free flight
loading", and "split-endcap loading", and the figures in which
these loading strategies are explained are given.
The fluorescence intensities are normalized to the fluorescence
intensity that is obtained using the conventional trap loading
strategy.
\begin{table}[htb!]
\caption{\label{tab:new:results} Comparison of experiments and
simulations using different trap loading strategies.}
\begin{tabular}{l|l|l|l|l|l}
\hline
Trap loading & Described  & Signal int. & Efficiency & $N$ & T \\
strategy     &  in Figure      & (Exp.)      & (Sim.)       & (Sim.)   & (Sim.) \\
\hline
Conventional & ~\ref{fig:new:old10kV}    & 1.0 & 15 \% & 1.0 & 59 mK\\
Free flight & ~\ref{fig:new:flatloading} & 4.0 & 27 \% & 1.8 & 51 mK\\
Split-endcap & ~\ref{fig:new:SECidea}    & 8.9 & 27 \% & 1.9 & 48 mK\\
\hline
\end{tabular}
\end{table}

It is seen that with the conventional quadrupole trap, the
efficiency of trap loading is increased by a factor 4 if a free
flight section is included in the loading sequence. Another factor
of 2.2 is gained if the left end-cap electrode is replaced by
split-endcap electrodes. It is noted that this value represents a
lower limit of the gain, as the present experiments are all
performed with the split-endcap electrodes in place. Compared to
the spherically symmetric endcap, that has a mere 4~mm
diameter opening, the slit offers a higher probability for the
molecules to enter the trap.

It is interesting to compare these experimental findings with trap
loading efficiencies that result from three dimensional trajectory
simulations. In general, it is difficult to obtain a quantitative
agreement between three dimensional trajectory simulations and
experimental results for trapping experiments. This is in part due to
the low velocity of the molecules during the trap loading process,
and hence their sensitivity to the exact details of the potentials
involved, and in part due to the complex shapes of the trap electrodes.
Nevertheless, these simulations are helpful to yield a qualitative
understanding of the trap loading process.

Three dimensional trajectory simulations have been performed for
all trap loading strategies. The calculated trap loading
efficiencies, defined as the fraction of the molecules that exit
the decelerator and that are still confined in the trap 20~ms
after the trap has been switched on, are presented in the fourth
column of Table \ref{tab:new:results}. For a direct comparison
with the experimental findings, however, the ratio between the
absolute number of molecules that are confined in the trap is of
relevance. This ratio can differ from the ratio of the
efficiencies, as the molecular packet has a different velocity in
the last stage of the decelerator for the different loading
strategies. The resulting number of molecules, normalized to the
number that is obtained using the "conventional" loading strategy,
is given in the fifth column of Table~\ref{tab:new:results}.
Finally, the temperature $T$ of the trapped sample of molecules
that follows from these simulations is given in the last column of
Table~\ref{tab:new:results}. This temperature is defined using the
average kinetic energy of the molecules $E_{kin}=3/2 k_B T$, where
$k_B$ is the Boltzmann constant.

It is seen that in the simulations, the performance of the
"conventional" trap loading strategy is worse than the "free
flight" or "split-endcap" loading strategy, in qualitative
agreement with the experimental findings. In contrast, the
simulations do not predict the improved loading efficiency for the
"split-endcap" compared to the "free flight" strategy that was
found experimentally (see table~\ref{tab:new:results}), and that
was expected from the one dimensional simulations presented in
figures \ref{fig:new:flatloading} and \ref{fig:new:SECidea}. This
somewhat surprising result indicates that the transverse focusing
properties of the split-endcap electrodes can diminish the gain
that is obtained in the longitudinal direction, as will be
discussed in detail below.

\section{Transverse motion during trap loading}\label{sec:transverse}

\begin{figure}[htb!]
   \centering
   \resizebox{\linewidth}{!}
   {\includegraphics[0,0][650,265]{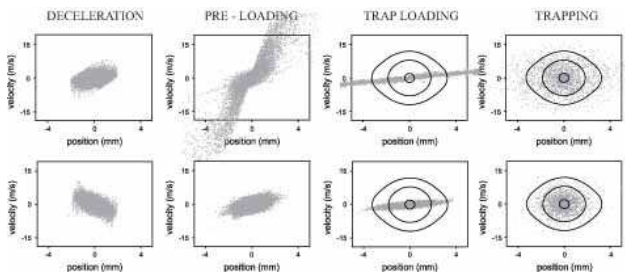}}
   \caption{Transverse ($y$ (top) and $z$ (bottom)) phase-space distributions at various stages of the experiment. These
   transverse phase-space snapshots belong to the longitudinal distributions presented in
   figure~\ref{fig:new:SECidea}.}
   \label{fig:new:sec-trans}
\end{figure}

Analogous to the evolution of the longitudinal phase-space
distribution during the trap loading process, the phase-space
distribution in both transverse directions can be derived from the
three dimensional trajectory simulations. These are shown for the
split-endcap loading strategy in figure~\ref{fig:new:sec-trans}.
The transverse coordinate $y$ (top panels) is defined along the
long axis of the last electrode pair. The transverse coordinate
$z$ (lower panels) is defined perpendicular to that axis. As can
be seen on the left side of figure~\ref{fig:new:sec-trans}, the
molecular packet has transverse dimensions of 4x4~mm$^2$ at the
exit of the decelerator, given by the distance between the
electrodes in each electrode pair. The split-endcap electrodes
face each other in the $y$ direction and create a pre-loading
field that is perpendicular to the field of the last deceleration
stage. Therefore, as the packet travels in the pre-loading field
it is not only decelerated in the longitudinal ($x$) direction,
but also gets strongly focused in the $y$ direction. The exact
strength of this focusing force depends on the longitudinal
position of the molecules, resulting in a non-uniform rotation of
the phase-space distribution. During the "pre-loading"
configuration, the molecules hardly experience focusing forces in
the $z$ direction (the direction along the slit of the
split-endcap electrodes), and in this direction the phase-space
distribution evolves more-or-less like in free flight. When the
packet progresses on the subsequent (cylindrically symmetric)
loading potential, the molecules get focused in the $y$ and $z$
direction equally strong. These focusing forces, however, are
limited, and in addition to a rotation in phase-space, the packet
also elongates spatially. The resulting phase-space distributions
at the moment the trap is switched into the "trapping"
configuration are shown in the third panel of
figure~\ref{fig:new:sec-trans}.

It is clear from these distributions that, in contrast to the
longitudinal phase-space overlap, the phase-space overlap with the
transverse trap acceptance is rather poor. In the $y$-direction,
the focusing force during the "pre-loading" configuration has been
too strong. A significant part of the molecules pass through a
focus, resulting in a large velocity spread when the packet enters
the "loading" potential. This large velocity spread is transferred
into a large position spread during the "loading" part of the
sequence. In the $z$-direction, however, the focusing forces are
too weak. The packet spreads out significantly, also resulting in
a large position spread at the time the trap is switched on.

\begin{figure}[htb!]
   \centering
   \resizebox{\linewidth}{!}
   {\includegraphics[0,0][650,537]{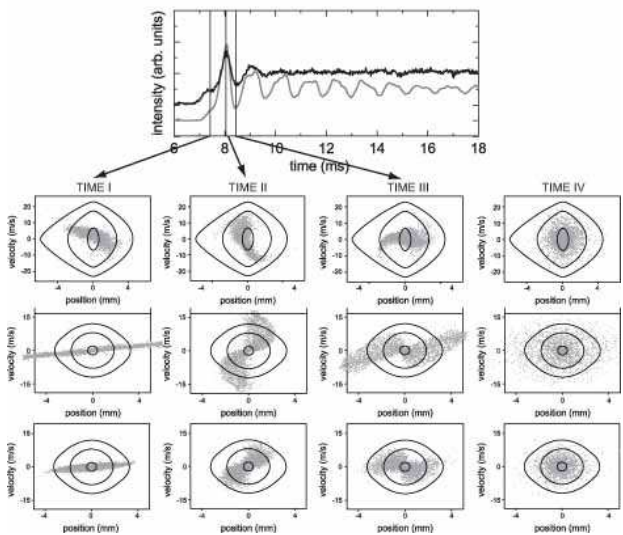}}
   \caption{Phase-space distributions in the longitudinal ($x$, top panel) and in
   both transversal ($y$, middle panel; $z$, lower panel) directions at different
   points in time, indicated with vertical lines in the time-of-flight
   profile shown on the top. Time~I
   is the time of switching from the loading to the trapping potential. Time~II and III correspond to the first maximum and
   minimum of the oscillations as observed in the time-of-flight profile of the trapping experiment, respectively. Time~IV
   is 20~ms after the trap has been switched on.}
   \label{fig:new:oscillations}
\end{figure}

The (transverse) mismatch between the phase-space distribution of
the molecular packet and the trap acceptance at the moment the
trapped is switched on can also be inferred from the
time-of-flight profile in figure~\ref{fig:new:tofflatloading}(d),
that is shown again in the upper curve of
figure~\ref{fig:new:oscillations}. Pronounced oscillations in the
signal intensity of the trapped molecules are observed in the
first milliseconds after the trap has been switched on. These
oscillations result from fluctuations in the density of molecules
within the detection volume, that can be observed if the detection
volume is considerably smaller than the total volume of the trap.
The detection volume is given by the 4~mm diameter of the
detection laser, that crosses both transverse coordinates $y$ and
$z$ under an angle of 45$^o$, and the opening angle of the
detection zone, given by the 6~mm diameter opening
in the right endcap. In figure~\ref{fig:new:oscillations} it is shown
that the fluctuations in density are mainly caused by the rotation
of the phase-space distribution of the packet in the transverse
directions.
At time~I, the time that the molecular packet arrives in the trap
center and the trap is switched on, the packet is longitudinally
and transversely (almost) velocity-focused; the packet is
spatially large (3x10x4~mm$^3$) and has a relative small velocity
spread (FWHM) of about 7, 5, and 2~m/s in the $x$, $y$, and $z$
direction, respectively. As can be seen from the top of the figure
this situation corresponds to a relative low LIF signal in the
time-of-flight profile, since a large fraction of the molecules is
located in the region of the trap that is not overlapped with the
detection laser. At time~II, the situation is exactly opposite.
The phase-space distributions have rotated and most molecules are
located in the center of the trap. The packet has a size of
1x2x2~mm$^3$ and a relatively high velocity spread of 21, 18, and
8~m/s in the $x$, $y$, and $z$ direction, respectively. This
situation corresponds with an intense LIF signal. The phase-space
distributions after 20~ms of trapping, shown in
figure~\ref{fig:new:oscillations} (time~IV), show that the packet
no longer has a clear structure in phase-space and the trap acceptance
is homogeneously filled.
The size of the molecular packet and the ratio between the
velocity distributions in the three directions are now given by
the shape of the phase-space acceptance of the trap. The spatial
distribution at this moment is 2x2x2~mm$^3$, the velocity
distribution 10, 8, and 7~m/s in the $x$, $y$, and $z$ direction,
respectively. This "steady-state" situation is reached as a result of
the coupling of the motion in all coordinates. The number of
oscillations that is actually seen in the simulated time-of-flight profile
critically depends on the details of the implementation of the detection
zone and the LIF collection optics in the  simulation, and differs from
the number of oscillations that is seen in the experiment.

\section{Conclusions}

In this paper, the efficiency of the loading of Stark-decelerated
molecular beams into electrostatic quadrupole traps has been
studied. These studies have been triggered by high losses that
have been observed during the trap loading process in previous
Stark deceleration and trapping experiments. These losses occur
because it is difficult to keep the molecular packet together in
the region between the end of the Stark decelerator and the first
electrode of the quadrupole trap. A new split-endcap quadrupole
trap, in which the cylindical symmetry of a quadrupole trap is
broken, is presented. This trap design allows for a continuation
of the sequence of potentials that are present inside the Stark
decelerator into the trap region, and effectively merges the exit
of the Stark decelerator with the electrostatic quadrupole trap.
The improved performance of this split-endcap quadrupole trap has
been experimentally verified by comparing the electrostatic
trapping of OH radicals using the new and conventional quadrupole
traps. Compared to the most successful loading strategy that has
been obtained in a quadrupole trap with the conventional electrode
design, an improvement in loading efficiency of a factor 2.2 has
experimentally been observed.

Three dimensional trajectory simulations of the trap loading
process reveal that in its current implementation, however, the
advantages of the split-endcap design are not yet fully exploited.
In the region between the Stark decelerator and the entrance of
the trap, the focusing of the molecular packet in one of the two
transverse directions is larger than desired, diminishing the gain
that is achieved in the longitudinal direction. The main lesson
learnt from these simulations is that the exact shape of the
split-endcap electrodes in the region of the slit is very critical
to the success of the split-endcap quadrupole trap. The slit
should be designed such that, prior to passing the slit, the
molecules experience a strong deceleration force but only modest
transverse focussing forces. This can, for instance, be
accomplished with split-endcap electrodes that are designed to
form an entrance opening slit with a more conical shape. For
future implementations of quadrupole traps that are based on the
split-endcap loading strategy, the critical interplay between
longitudinal and transverse forces during trap loading needs to be
thoroughly investigated using three dimensional trajectory
simulations.


\end{document}